\RequirePackage{fix-cm}
\documentclass[pdftex,twocolumn,epjc3upd,final]{svjour3} 

\RequirePackage[T1]{fontenc}
\smartqed  % flush right qed marks, e.g. at end of proof
\RequirePackage{graphicx}
\RequirePackage{mathptmx}
\RequirePackage[numbers,sort&compress]{natbib}
\RequirePackage[colorlinks,citecolor=blue,urlcolor=blue,linkcolor=blue]{hyperref}
\usepackage{amssymb}
\usepackage[usenames,dvipsnames]{xcolor}
\usepackage[normalem]{ulem}
\usepackage[dvipsnames]{xcolor}
\usepackage[symbol]{footmisc}

\journalname{Eur. Phys. J. C}
\usepackage{graphicx}           % \includegraphics with arguments
\usepackage{array}              % More table column format options
\usepackage{placeins}           % \FloatBarrier
\usepackage{amsmath}            % Vectors and much more
\usepackage{amssymb}            % lesssim, gtrsim, etc
\usepackage[version=4]{mhchem}  % Chemical element symbols
\usepackage{siunitx}            % \SI and \si for units
\usepackage{xspace}             % Spaces that don't get eaten

\usepackage{verbatim}           % comment environment
\usepackage{pgf}                % Package defining color names (and some other stuff I don't use)
\usepackage[utf8]{inputenc}     % Unicode symbols in latex source
\usepackage{microtype}          % "subliminal enhancements"

\usepackage[switch, displaymath, mathlines]{lineno}

\newcommand\hrefarxiv[1]{\href{http://arxiv.org/abs/#1}{arXiv:#1}}

\begin{document}

%\linenumbers

\title{Solar Neutrino Detection Sensitivity in DARWIN via Electron Scattering}


\author{
J.~Aalbers\thanksref{stockholm} \and
F.~Agostini\thanksref{bologna} \and 
S.~E.~M.~Ahmed~Maouloud\thanksref{paris} \and 
M.~Alfonsi\thanksref{mainz} \and
L.~Althueser\thanksref{munster} \and
F.~D.~Amaro\thanksref{coimbra} \and
J.~Angevaare\thanksref{nikhef} \and
V.~C.~Antochi\thanksref{stockholm} \and
B.~Antunovic\thanksref{vinca}\textsuperscript{,}\footnotemark[2] \and
E.~Aprile\thanksref{columbia}
% ...
(DARWIN Collaboration\thanksref{e3})
}

\thankstext{e1}{shayne@physik.uzh.ch}
\thankstext{e3}{darwin@lngs.infn.it}

\institute{Oskar Klein Centre, Department of Physics, Stockholm University, AlbaNova, Stockholm SE-10691, Sweden \label{stockholm} \and 
Department of Physics and Astronomy, University of Bologna and INFN-Bologna, 40126 Bologna, Italy \label{bologna} \and 
LPNHE, Universit\'{e} Pierre et Marie Curie, Universit\'{e} Paris Diderot, CNRS/IN2P3, Paris 75252, France \label{paris} \and
Institut f\"ur Physik \& Exzellenzcluster  PRISMA$^{+}$, Johannes Gutenberg-Universit\"at Mainz, 55099 Mainz, Germany \label{mainz} \and
Institut f\"ur Kernphysik, Westf\"alische Wilhelms-Universit\"at M\"unster, 48149 M\"unster, Germany \label{munster} \and
LIBPhys, Department of Physics, University of Coimbra, 3004-516 Coimbra, Portugal \label{coimbra} \and 
Nikhef and the University of Amsterdam, Science Park, 1098XG Amsterdam, Netherlands \label{nikhef}  \and
Vinca Institute of Nuclear Science, University of Belgrade, Mihajla Petrovica Alasa 12-14. Belgrade, Serbia \label{vinca} \and
Physics Department, Columbia University, New York, NY 10027, USA \label{columbia}
% ...
}

\date{Received: date / Accepted: date}

\maketitle

\sloppy

\footnotetext[2]{Also at University of Banja Luka, Bosnia and Herzegovina}
\footnotetext[3]{Also at Institute for Subatomic Physics, Utrecht University, Utrecht, Netherlands}
\footnotetext[4]{Wallenberg Academy Fellow} 
\footnotetext[5]{Also at Coimbra Polytechnic - ISEC, Coimbra, Portugal} 
\footnotetext[6]{Also at Faculty of Mathematics}
\footnotetext[7]{Also at SISSA, Data Science Excellence Department, Trieste, Italy}

\normalsize{
\begin{abstract}
We detail the sensitivity of the proposed liquid xenon DARWIN observatory to solar neutrinos via elastic electron scattering. We find that DARWIN will have the potential to measure the fluxes of five solar neutrino components: $pp$, $^7$Be, $^{13}$N, $^{15}$O and $pep$. The precision of the $^{13}$N, $^{15}$O and $pep$ components is hindered by the double-beta decay of $^{136}$Xe and, thus, would benefit from a depleted target. A high-statistics observation of $pp$ neutrinos would allow us to infer the values of the electroweak mixing angle, $\sin^2\theta_w$, and the electron-type neutrino survival probability, $P_{ee}$, in the electron recoil energy region from a few keV up to 200\,keV for the first time, with relative precision of 5\% and 4\%, respectively, with 10 live years of data and a 30\,tonne fiducial volume. An observation of $pp$ and $^7$Be neutrinos would constrain the neutrino-inferred solar luminosity down to 0.2\%. A combination of all flux measurements would distinguish between the high- (GS98) and low-metallicity (AGS09) solar models with 2.1-2.5$\sigma$ significance, independent of external measurements from other experiments or a measurement of $^8$B neutrinos through coherent elastic neutrino-nucleus scattering in DARWIN. Finally, we demonstrate that with a depleted target DARWIN may be sensitive to the neutrino capture process of $^{131}$Xe.
\end{abstract}
}

\section{Introduction}
Current and future liquid xenon (LXe) direct detection dark matter experiments, such as XENONnT~\cite{Aprile:2020vtw}, LZ~\cite{Akerib:2018dfk}, and DARWIN~\cite{Aalbers:2016jon}, will exhibit sensitivity to neutrinos at the $\sim$MeV scale. Typically, neutrinos have been regarded as backgrounds in the search for dark matter (DM)~\cite{Strigari:2009bq,Billard:2013qya}; but, as signals, they present opportunities to characterize their sources and pursue physics beyond the Standard Model (SM)~\cite{Lang:2016zhv,Baudis:2013qla,Newstead:2018muu,Dutta:2017nht,Cerdeno:2016sfi,AristizabalSierra:2017joc,Gonzalez-Garcia:2018dep}. While DM remains the primary objective, detectors with multi-tonne (t) xenon targets will seek neutrino signals without the need for additional investment.

Solar neutrinos, in particular, are observable in dark matter detectors through two types of interactions: elastic electron scattering (ES) and coherent elastic neutrino-nucleus scattering (CEvNS)~\cite{Billard:2014yka,Akimov:2017ade}. In the SM, ES is mediated by both charged and neutral currents. The former is only possible for $\nu_e$, which creates nearly an order of magnitude of difference between the interaction rates of $\nu_e$ and $\nu_{\mu,\tau}$. In total, the cross section for ES is $\sim10^{-43}$\,cm$^2$. On the other hand, CEvNS is mediated only by neutral current, with a cross section of $\sim10^{-39}$\,cm$^2$ that is strongly determined by the target's neutron number. With their different sensitivities to the components of the solar neutrino flux, these two channels provide complementarity over a wide range of neutrino energies.

Dedicated solar neutrino experiments have made numerous observations of ES with water, heavy water, and liquid scintillator targets. Borexino independently measured the fluxes of the lower-energy $pp$~\cite{Bellini:2014uqa}, $^7$Be~\cite{Bellini:2011rx}, and $pep$~\cite{Bellini:2011nga} components. Subsequently, Borexino presented the first results from simultaneous spectroscopy of these three components above 0.19\,MeV, yielding the most precise measurements to date~\cite{Agostini:2017ixy,Agostini:2018glp}. Most recently, Borexino reported the first direct observation of carbon, nitrogen and oxygen (CNO) neutrinos at 5$\sigma$~\cite{Agostini:2020mfq}. Five experiments, Borexino~\cite{Bellini:2008mr,Agostini:2017cav}, Super-K~\cite{Abe:2010hy}, KamLAND~\cite{Abe:2011em}, SNO~\cite{Aharmim:2011vm}, and SNO+~\cite{Anderson:2018ukb}, have measured the higher-energy $^8$B flux. COHERENT made the first observation of CEvNS~\cite{cevns}, but astrophysical neutrinos have yet to be detected in this way.

After decades of investigation, important questions about the Sun persist. From an astrophysical perspective, the most salient issue lies in the solar abundance problem. The more recent low-metallicity (low-Z) AGSS09 SSM~\cite{Serenelli:2016dgz,Asplund:2009ags} would seem to better represent the photosphere than its predecessor, the high-metallicity (high-Z) GS98 SSM~\cite{Serenelli:2016dgz,Grevesse:1998gs}. However, a comparison of individual flux measurements with theoretical predictions tends to favor the high-Z SSM, contradicting the common assumption that abundances in the radiative envelope are the same as those in the photosphere. This preference is further supported by helioseismic data that have long since disfavored a low-Z model~\cite{Serenelli:2009yc}. As carbon, nitrogen and oxygen constitute the majority of heavy elements in the Sun, their neutrino fluxes are the most sensitive to metallicity. A combined analysis of available measurements remains inconclusive, but a relative uncertainty of $\sim$15\% on a combined CNO flux measurement would begin to favor one model over another~\cite{Bergstrom:2016cbh,Agostini:2018glp,Agostini:2020lci}. While less sensitive to metallicity than CNO neutrinos, an improved measurement of the $^8$B flux would also help to distinguish them further.

Measurements of electroweak and oscillation parameters play an important role in the understanding of the SM and the search for new physics~\cite{Kumar:2013yoa,Maltoni:2015kca}. Non-standard neutrino interactions (NSI) might modify the Large Mixing Angle (MSW-LMA) solution to the solar neutrino problem. Solar neutrinos serve as one probe to observe or set bounds on NSI. Two of these parameters, the electroweak mixing angle ($\sin^2\theta_w$) and the $\nu_e$ survival probability ($P_{ee}$), which is the probability that a $\nu_e$ is detected on Earth as a $\nu_e$, may be measured with the ES process. The survival probability is determined by the elements of the unitary lepton mixing matrix. Only the higher energy solar neutrinos, $^8$B and $hep$, experience significant matter oscillation (MSW) effects. On one hand, atomic parity violation in cesium at 2.4\,MeV yields the lowest energy at which $\sin^2\theta_w$ has been measured~\cite{Davoudiasl:2015bua}. On the other, Borexino has provided the lowest-energy measurement of the $\nu_e$ survival probability extracted from the tail of the proton-proton distribution ($>$0.19\,MeV)~\cite{Agostini:2018glp}. No experiment, thus far, has been able to access energies below these respective thresholds. DARWIN will measure $\sin^2\theta_w$ and $P_{ee}$ for the first time in the (recoil) energy region [1,200]\,keV.

In this manuscript, the authors highlight the efficacy with which DARWIN will shed light on solar and neutrino physics through elastic electron scattering. Details about the sensitivity to each component of the solar neutrino flux are provided first. Then, the precision with which DARWIN would reconstruct $\sin^2\theta_w$ and $P_{ee}$ in the low energy range [1,200]\,keV is illustrated. Lastly, the extent to which a combined analysis of neutrino flux measurements would resolve the solar abundance problem is demonstrated.

\section{The DARWIN Experiment}
The DARWIN observatory, which may begin operation as early as 2026, is a next-generation dark matter experiment that will operate with 50\,t (40\,t active) of xenon in a cylindrical, dual-phase time projection chamber (TPC) that is 2.6\,m in both height and diameter~\cite{Aalbers:2016jon}. The TPC will be placed underground in a double-walled cryostat vessel shielded by water Cherenkov and neutron vetoes that enable one to observe cosmogenic muons and their progeny. The TPC will be equipped to read out both light and charge signals. The location of the DARWIN observatory has not yet been determined. Discussions of background presume the overburden at the Laboratori Nazionali del Gran Sasso (3500 meter water equivalent).

A charged particle that interacts in liquid xenon produces photons (scintillation) and electrons (ionization). The TPC promptly detects these photons as a primary ``S1" scintillation signal with photosensors (PMTs, SiPMs, or an alternative) instrumented in arrays at the top and bottom of the target region. An electric field, $\mathcal{O}(0.1)$\,kV/cm, applied in the liquid volume drifts the electrons upward, and then a second field, $\mathcal{O}(1)$\,kV/cm, extracts them into the gas phase, where electroluminescence generates an amplified ``S2" scintillation signal. The radial position of an interaction is reconstructed with the S2 light pattern in the top array, while its depth is inferred from the time delay between S1 and S2, up to $\mathcal{O}$(1\,ms). Together, S1 and S2 reconstruct the energy of the event with excellent resolution. The charge-to-light ratio S2/S1 discriminates between scatters off electrons and those off nuclei~\cite{Lebedenko:2008gb}. The combination of position, energy and discrimination allows for strong event selections to mitigate sources of background.

\begin{figure*}
\centering
\begin{tabular}{l}
\includegraphics[width=2\columnwidth]{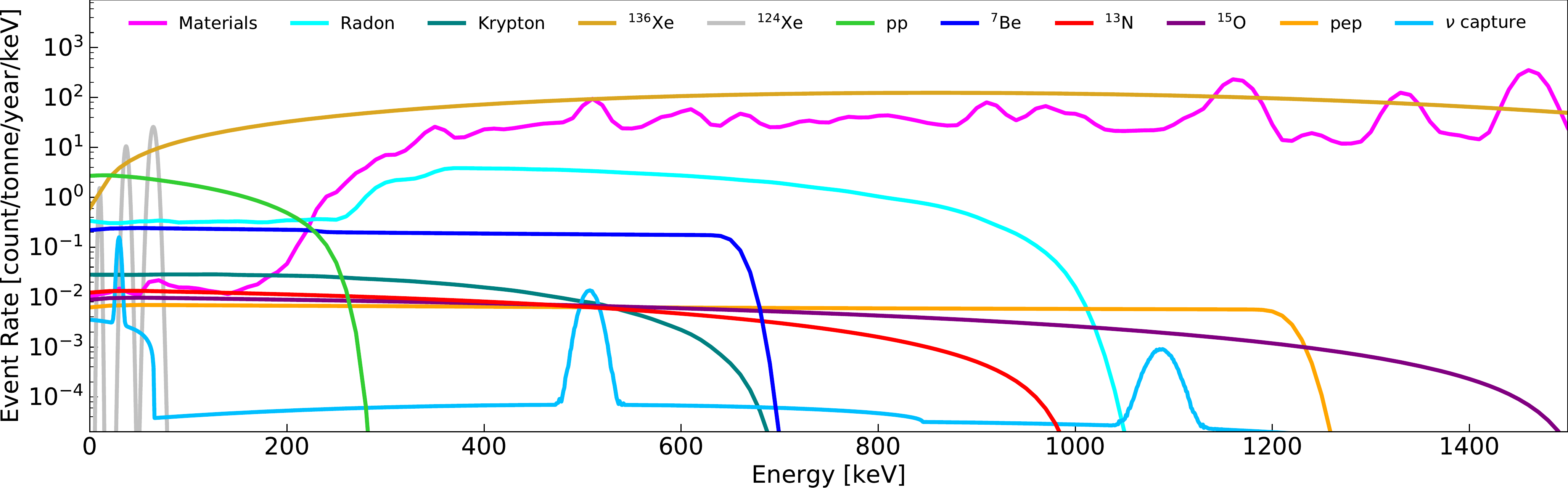}
\end{tabular}
\caption{The electron recoil spectra of five solar neutrino components, neutrino capture on $^{131}$Xe, and five backgrounds up to 1.5\,MeV. The solar components follow from the high-Z SSM model. The materials and $^{136}$Xe events in [1.5,3]\,MeV (not shown) are also used in the statistical analysis. The materials component is based on a selection of events in a 30\,t fiducial volume.}
\label{fig:spectra}
\end{figure*}

The most troublesome background for a solar neutrino search (ES) arises from the $^{222}$Rn emanated by detector components. More precisely, the $^{214}$Pb daughter decays directly to $^{214}$Bi with a branching ratio of 11\%, emitting a lone $\beta$ with an energy up to $Q=1.02$\,MeV~\cite{NuclearData}. Otherwise, $^{214}$Pb decays to an excited state of $^{214}$Bi that emits a $\gamma$ coincident with the $\beta$ to create a sharp rise above the lone-$\beta$ continuum. There are several excited states that contribute, starting at 0.274\,MeV, as illustrated in Figure~\ref{fig:spectra}. With a long half-life (3.8\,d), $^{222}$Rn distributes itself homogeneously in the LXe volume, such that it is not reduced with the deliberate selection of an inner volume, known as fiducialization. As radon mitigation is vital for the dark matter search, efforts are already underway to achieve a lower $^{222}$Rn concentration. The detector materials in DARWIN will be screened for low radon emanation through a dedicated radioassay program, as in XENON1T/nT~\cite{Aprile:2017ilq}, to minimize the number of events in the low-energy region of interest. Gamma-ray spectroscopy and inductively-couple plasma mass spectrometry will provide systematic measurements of a wide variety of raw materials. The known activities of impurities will allow for the careful selection of these materials and the deliberate placement of fabricated components during construction. Furthermore, DARWIN is assumed to have a cryogenic radon distillation column integrated into the (gas) purification loop in order to counteract the continuous generation of $^{222}$Rn from trace amounts of $^{226}$Ra. Radon accumulates in a reboiler at the bottom of the distillation column and remains there until disintegration. Meanwhile, the xenon is purified and extracted to the top, where it is reintroduced into the system. Cryogenic distillation has been successfully applied in XENON100~\cite{Aprile:2017kop} and XENON1T~\cite{thesis:Murra}. In the latter case, a reduction of $\sim$20\% was achieved. For this study, a target $^{222}$Rn activity of 0.1\,$\mu$Bq/kg is assumed.

A second background comes from intrinsic $^{85}$Kr, a $\beta$ emitter ($Q=0.687$\,MeV; T$_{1/2}=10.8$\,y) that remains in the xenon volume after extraction from the atmosphere. As with $^{222}$Rn, $^{85}$Kr homogeneously distributes itself in the LXe volume. A krypton distillation column, designed to separate krypton and xenon through their different vapor pressures, may be deployed prior to a science run in order to reduce the krypton concentration to the desired level. The higher volatility of krypton in a static LXe reservoir leads to krypton enrichment in a gaseous xenon phase above. A portion of this enriched gas is removed at the top of the distillation column, and the purified xenon is extracted through the bottom. A series of distillation stages is needed to reach a separation factor greater than $10^5$. XENON1T has already demonstrated a concentration $^\text{nat}$Kr/Xe $<$ 360\,ppq~\cite{Aprile:2019dme}. Krypton may be further reduced at any time via online distillation, as applied in XENON1T~\cite{thesis:Murra,Aprile:2016xhi}. A concentration of 2\,ppq is assumed in this study, but--even at its current level--it has a negligible effect on measurements of flux, the electroweak mixing angle, and the electron-type survival probability. Only an observation of neutrino capture would require a $\mathcal{O}\text{(ppq)}$ concentration.

Long-lived radionuclides in detector materials constitute a third class of background events. The decay chains of $^{238}$U, $^{232}$Th and $^{235}$U generate various $\alpha$ and $\beta$ particles as well as $\gamma$ rays. The main contributors of $\gamma$ rays from these three chains are $^{214}$Bi (2.45\,MeV) and $^{208}$Tl (2.61\,MeV), including the background induced by decays in the non-instrumented xenon volume around the TPC. Additional $\gamma$ rays are emitted in the decays of $^{137}$Cs (0.662\,MeV), $^{40}$K (1.46\,MeV), and the daughters of $^{60}$Co (1.17 and 1.33\,MeV) and $^{44}$Ti (2.66\,MeV). The $\alpha$ and $\beta$ particles do not travel far and thus are eliminated with fiducialization. However, the $\gamma$ rays, with attenuation lengths up to $\sim$6\,cm at 1\,MeV, penetrate the innermost region, where they experience photoabsorption or Compton scattering. The more notable contributors have historically been the stainless steel cryostat and photosensors~\cite{Aprile:2017ilq}. A materials background component derived from the DARWIN simulation in~\cite{Agostini:2020adk}, which considers a more radiopure titanium cryostat, is included. The long-lived (T$_{1/2}=59.1$\,y), cosmogenically activated $^{44}$Ti contributes to the materials background through its daughter $^{44}$Sc, which subsequently decays (T$_{1/2}=3.8$\,h) with the emission of a 2.66\,MeV gamma. The simulated materials spectrum is adapted to this study by incorporating position-dependent multiscatter resolution, 3-15\,mm, and selecting events within a 30\,t super-ellipsoidal fiducial volume that minimizes the contribution of these Compton scatters below recoil energies of $\sim$200\,keV. With 10 live years of data, DARWIN would accrue 300 tonne-years (ty) of exposure, compared to the 200\,ty goal for the dark matter search.

Finally, unstable xenon isotopes pose a potential background in the search for ES of solar neutrinos. The isotope $^{136}$Xe, which occurs naturally with an abundance of 8.9\%, undergoes double-beta decay ($Q=2.46$\,MeV; T$_{1/2}=2.17\cdot10^{21}$\,y). The resultant spectrum circumscribes the entire signal region of interest. Furthermore, the cosmic-muon-induced neutron capture process of $^{136}$Xe creates $^{137}$Xe, which then beta decays ($Q=4.16$\,MeV; T$_{1/2}=3.82$\,min). The impact of $^{137}$Xe proves to be negligible at the level of 10$^{-3}$ per tonne-year per keV, three orders of magnitude lower than $^{136}$Xe double-beta decay~\cite{Agostini:2020adk}. These $^{136}$Xe background contributions are removable through isotopic depletion. The neutrinoless double-beta decay experiment EXO-200~\cite{Anton:2019wmi}, which uses a LXe volume enriched in $^{136}$Xe to 80.6\%, has already demonstrated the possibility to alter the isotopic composition through ultracentrifugation~\cite{Auger:2012gs}. Depletion, however, would diminish the prospects for a neutrinoless double-beta decay search with $^{136}$Xe in DARWIN~\cite{Agostini:2020adk}. Lastly, $^{124}$Xe decays via double electron capture (T$_{1/2}=1.4\cdot10^{22}$\,y)~\cite{Doi:1992,Wittweg:2020fak}, as first observed in XENON1T~\cite{XENON:2019dti}. The subsequent cascade of Auger electrons and X-rays is observed as a single peak at 64.3 (36.7; 9.8)\,keV with a branching ratio of 0.75 (0.23; 0.017), following the fast atomic process and their sub-millimeter spread in liquid xenon. With an abundance of 0.1\%, one expects a total of 228 double electron capture events per tonne-year.

%% Results figure

%\begin{figure}[!t]
%\centering
%\includegraphics[width=\columnwidth]{energy_spectra.pdf}
%\includegraphics[width=\columnwidth]{energy_spectra_zoom_350.pdf}
%\caption{(top) The electron recoil spectra of five solar neutrino components, neutrino %capture on $^{131}$Xe (solid), and five backgrounds (dashed) up to 1.5\,MeV. The solar %components follow from the high-Z SSM model. The materials and $^{136}$Xe events in %[1.5,3]\,MeV (not shown) are also used in the statistical analysis. The materials %component includes events in a 30\,t fiducial volume. (bottom) A zoomed spectrum of the %lower energy components.}
%\label{fig:spectra}
%\end{figure}

\section{Solar Neutrinos in DARWIN} \label{sec:nueventrates}

DARWIN will be optimized for the detection of low-energy nuclear recoils.  This fact also implies that DARWIN will be well equipped to detect ES with high efficiency and excellent energy resolution. In the following, the expected event rates for the individual solar components are calculated.

The spectral fluxes of $pp$, $^{13}$N, and $^{15}$O neutrinos are represented with the $\beta$ form,
\begin{linenomath}\begin{equation}
\frac{d\Phi_i}{dE_\nu}=\Phi_iA(x_i-E_\nu)[(x_i-E_\nu)^2-m_e^2]^{\frac{1}{2}}E_\nu^2 ,
\end{equation}\end{linenomath}
where $x_i\equiv Q_i+m_e$, $Q_i$ and  $\Phi_i$ are the characteristic maximal energy and the flux scale of neutrino component $i$, respectively, $m_e$ is the electron mass, $A$ is the corresponding normalization factor, and $E_\nu$ is the energy of the emitted neutrino. In contrast, $^{7}$Be and $pep$ neutrinos are monoenergetic. The $^{7}$Be neutrinos are emitted at 0.862\,MeV (0.384\,MeV) with a branching ratio of 90\% (10\%), while the $pep$ neutrinos have an energy of 1.44\,MeV. The flux scales are taken from the high-metallicity solar model~\cite{Vinyoles:2016djt}.

These spectral fluxes are convolved with the differential cross section of elastic electron-neutrino scattering:
\begin{linenomath}\begin{equation}
\frac{dR_i}{dE_r}=N_e \sum_j \int P_{ej} \frac{d\Phi_i}{dE_\nu}\frac{d\sigma_j}{dE_r}dE_\nu ,
\end{equation}\end{linenomath}
where $P_{ej}$ is the oscillation probability of lepton flavor $j$ to the electron neutrino, $N_e=2.48\times10^{29}$ is the number of target electrons per tonne of xenon, and $E_r$ is the energy of the induced recoil. The flux scales, maximum neutrino energies and survival probabilities are listed in Table~\ref{table:solpars}. The values of survival probability follow the MSW-LMA solution at low energies in the vacuum-dominated regime~\cite{Capozzi:2017ipn}. The differential cross section is given by~\cite{Marciano:2003eq,Formaggio:2013kya}
\begin{linenomath}\begin{equation}
\frac{d\sigma}{dE_r}=\frac{2G_F^2m_e}{\pi}\bigg[g_L^2+g_R^2\bigg(1-\frac{E_r}{E_\nu}\bigg)^2-g_Lg_R\frac{m_eE_r}{E_\nu^2}\bigg] ,
\end{equation}\end{linenomath}
with the coupling parameters $g_L=\sin^2\theta_w-\frac{1}{2}$ and $g_R=\sin^2\theta_w.$
For the $\nu_e$, $g_L\rightarrow g_L+1$ to account for its charged current interactions. A value of $\sin^2\theta_w=0.2387$~\cite{Erler:2004in} is assumed. In order to induce an electronic recoil, an incident neutrino must possess more energy than the binding energy of a given shell; and, when a recoil occurs, its energy is lowered accordingly. For this reason, xenon is not completely sensitive to neutrinos with the lowest energies. This effect is incorporated in the neutrino scattering rates with a step function defined by the series of discrete electron binding energies from 12\,eV to 35\,keV. This ultimately leads to a suppression of a few percent in the $pp$ neutrino event rate and negligible reductions for the other solar neutrino components. The Gaussian energy resolution obtained in XENON1T~\cite{Aprile:2020yad}, which remain unchanged with the step approximation, is also applied:
\begin{linenomath}\begin{equation}
\frac{\sigma(E_r)}{E_r}=\frac{0.3171}{\sqrt{E_r \text{[keV]}}}+0.0015 .
\end{equation}\end{linenomath}

\begin{table}[!tbp]
\centering
\begin{tabular}{c c c c c}
\hline%\hline
component & \: $\Phi [\text{cm}^{-2}\text{s}^{-1}]$ \: & \: $\sigma$ [\%] \: & \: $Q [\text{keV}]$ \: & \: $P_{ee}$ \: \\\hline
pp        & 5.98$\cdot$10$^{10}$ & 0.6 & 420 & 0.55 \\
$^7$Be    & 4.93$\cdot$10$^{9}$ & 6  & 862, 384 & 0.52 \\
$^{13}$N  & 2.78$\cdot$10$^{8}$ & 15 & 1200 & 0.52 \\
$^{15}$O  & 2.05$\cdot$10$^{8}$ & 18 & 1732 & 0.50 \\
pep       & 1.44$\cdot$10$^{8}$ & 1 & 1442 & 0.50 \\
%$^{17}$F  & 5.29$\cdot$10$^{6}$  & 1732 & 0.50 \\
%$^{8}$B   & 5.46$\cdot$10$^{6}$  & 16360 & 0.47 \\
%hep       & 7.98$\cdot$10$^{3}$  & 18773 & 0.47 \\
\hline%\hline
\end{tabular}
\caption{The characteristic values of the flux scales~\cite{Vinyoles:2016djt}, their relative uncertainties, the maximum neutrino energies, and the MSW-LMA $\nu_e$ survival probability~\cite{Capozzi:2017ipn} used in this study.\label{table:solpars}}
\end{table}

The $pp$ neutrinos constitute the most prominent component due to the low energy threshold achievable in LXe TPCs. Here, a threshold of 1\,keV is assumed, yielding an integrated rate of 365 events per tonne-year. This high rate presents an opportunity to probe $\sin^2\theta_w$ for the first time below $\sim$200\,keV, to improve upon the precision of existing measurements of $P_{ee}$ at low energies, and to further constrain the neutrino-inferred measurement of solar luminosity.

The $^7$Be neutrinos comprise the second most prominent component. The larger branch contributes 133 events per tonne-year, while the smaller one contributes 7.6 events. The $^7$Be flux is more sensitive to solar metallicity and, as such, it may be combined with a high-precision measurement of the $pp$ flux to make an initial assessment of different metallicity models.

The third most prominent components are those of $^{13}$N, $^{15}$O, and $pep$, which induce 6.5, 7.1 and 7.6 events per tonne-year, respectively. Despite having the lowest rate, $^{13}$N events fall within a narrower energy range than either $^{15}$O or $pep$, such that the $^{13}$N spectrum rises above both below $\sim$0.4\,MeV. Consequently, it is possible for DARWIN to make a statistically significant observation of CNO neutrinos by exploiting higher statistics at lower energies. As the most sensitive to metallicity, being 30\% higher in the high-Z scenario, measurements of the $^{13}$N and $^{15}$O fluxes would greatly enhance the capability to distinguish between solar models. The rates of $^{17}$F, $^8$B and $hep$ neutrinos are negligible.

Finally, the possibility of neutrino capture on $^{131}$Xe ($Q=0.355$\,MeV) is considered. This isotope is the only one with a sufficiently low Q-value to exhibit sensitivity to solar neutrinos through this type of charged-current interaction. The expected observable signature consists of two signals: a prompt electron and a combination of X-rays and Auger electrons that are emitted together in the subsequent electron capture (EC) decay of $^{131}$Cs$^+$  (T$_{1/2}=9.69$\,days). The prompt electrons would create a spectrum that mirrors those of the spectral neutrino fluxes shifted to lower energies by the $Q$-value of this reaction ($E_e = E_\nu-Q$). The EC decay would appear as a Gaussian peak at 0.030\,MeV. The long half-life of the EC process precludes delayed coincidence of these two signatures. The contribution of each solar component (including $^8$B) follows from~\cite{Georgadze:1997vcp}. There are three distinct peaks visible in Figure 1. The two higher energy peaks come from capturing the monoenergetic $^7$Be and $pep$ neutrinos; while the feature below 50\,keV is a combination of EC, the lower branch of $^7$Be, and the tail of the $pp$ spectrum. In contrast to $pp$ neutrinos, the EC peak of the neutrino capture process provides a probe of the electron-type survival probability that depends on neutrinos in both the vaccuum- and matter-dominated regimes. As the five neutrino flux components considered here can be measured more precisely than the neutrino capture process, one may use the EC peak to infer the survival probability at high neutrino energies, as well, with an uncertainty comparable to current measurements. Accounting for both the survival probability and the $^{131}$Xe abundance of 21.2\%, one expects 1.23 neutrino capture events per tonne-year with a natural xenon target.

\section{Flux and Luminosity} \label{sec:fluxsen}
Having defined the signal and background models, an assessment of DARWIN's sensitivity to each of the neutrino components is made. A multivariate spectral fit of all 11 components up to 3\,MeV is employed. Each signal and background component is represented by a single scale parameter in the fit. These scale parameters define the set of maximum likelihood estimators for the neutrino flux components, $\vec{f}=\{f_{pp},f_{Be},f_N,f_O,f_{pep}\}$, and the neutrino capture rate, $f_\text{cap}$, given the electroweak and oscillation parameters, $\vec{\theta}=\{\sin^2\theta_w,P_{ee}\}$:
\begin{linenomath}\begin{equation}
P(n_j | \mu_j(\vec{f}, f_\text{cap}) ) = \mathcal{L}(\vec{f}, f_\text{cap}) = \prod_{j=1} \frac{\mu_j^{n_j}}{n_j ! }e^{-\mu_j} .
\end{equation}\end{linenomath}
These two parameters, which are fixed to the values given in Section~\ref{sec:nueventrates} for this low energy region, predict the average number of events in the $j$th energy bin, $\mu_j$, while the observed number of events in that bin, $n_j$, is randomly sampled. The background-only region, [1.5,3]\,MeV, is also used to constrain the uncertainties in the normalization of the materials and $^{136}$Xe backgrounds at lower energies. The key advantage of the full spectral fit compared to a counting experiment in the low energy region of interest is that the constraints on the larger backgrounds, from $^{136}$Xe and materials, may be improved from $\mathcal{O}(10)$\% to $\mathcal{O}(0.1)$\% or better. The scale of each signal and background component is left completely free throughout the fitting routine.

\begin{figure}[!t]
\centering
\includegraphics[width=1\columnwidth]{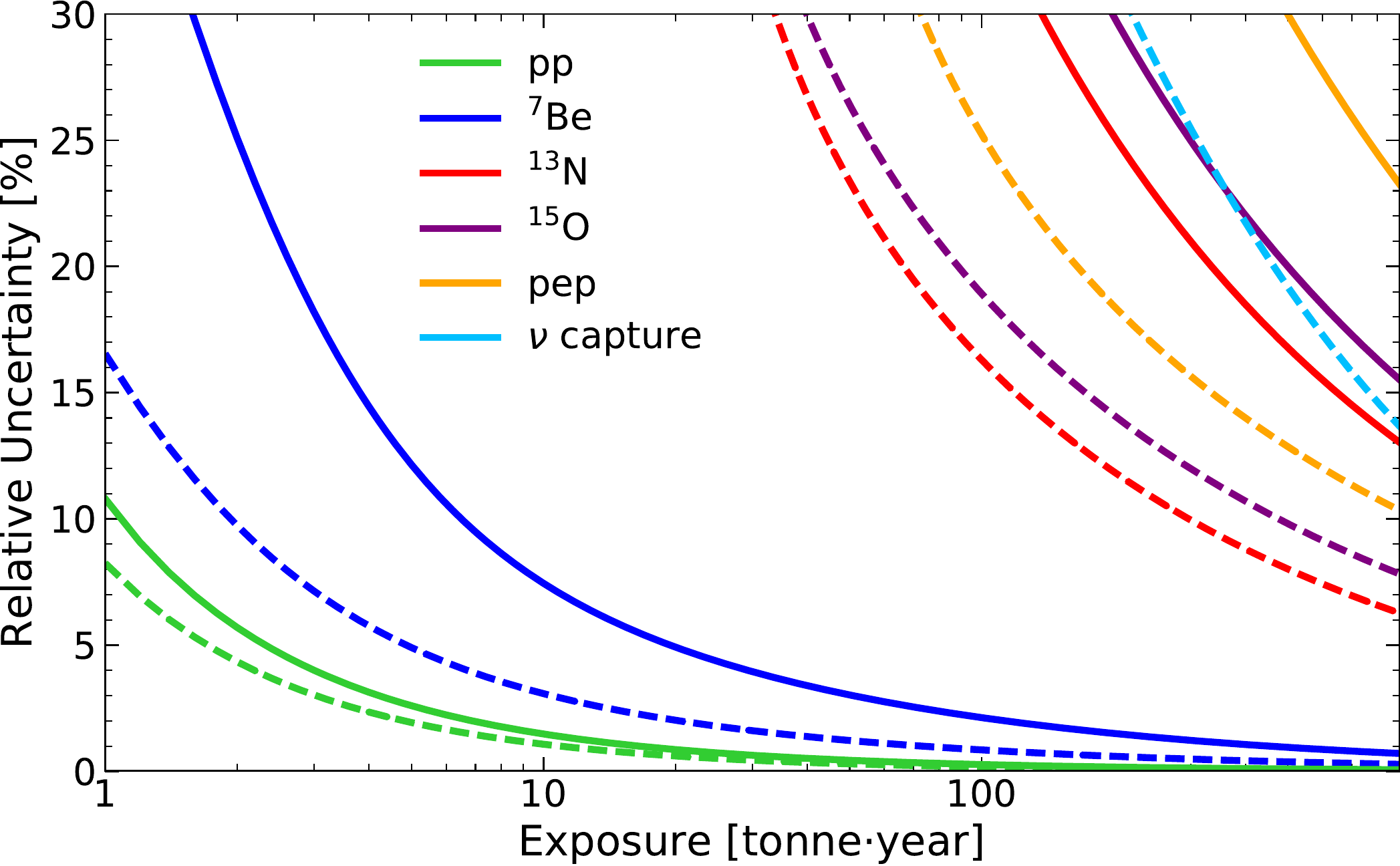}
\includegraphics[width=1\columnwidth]{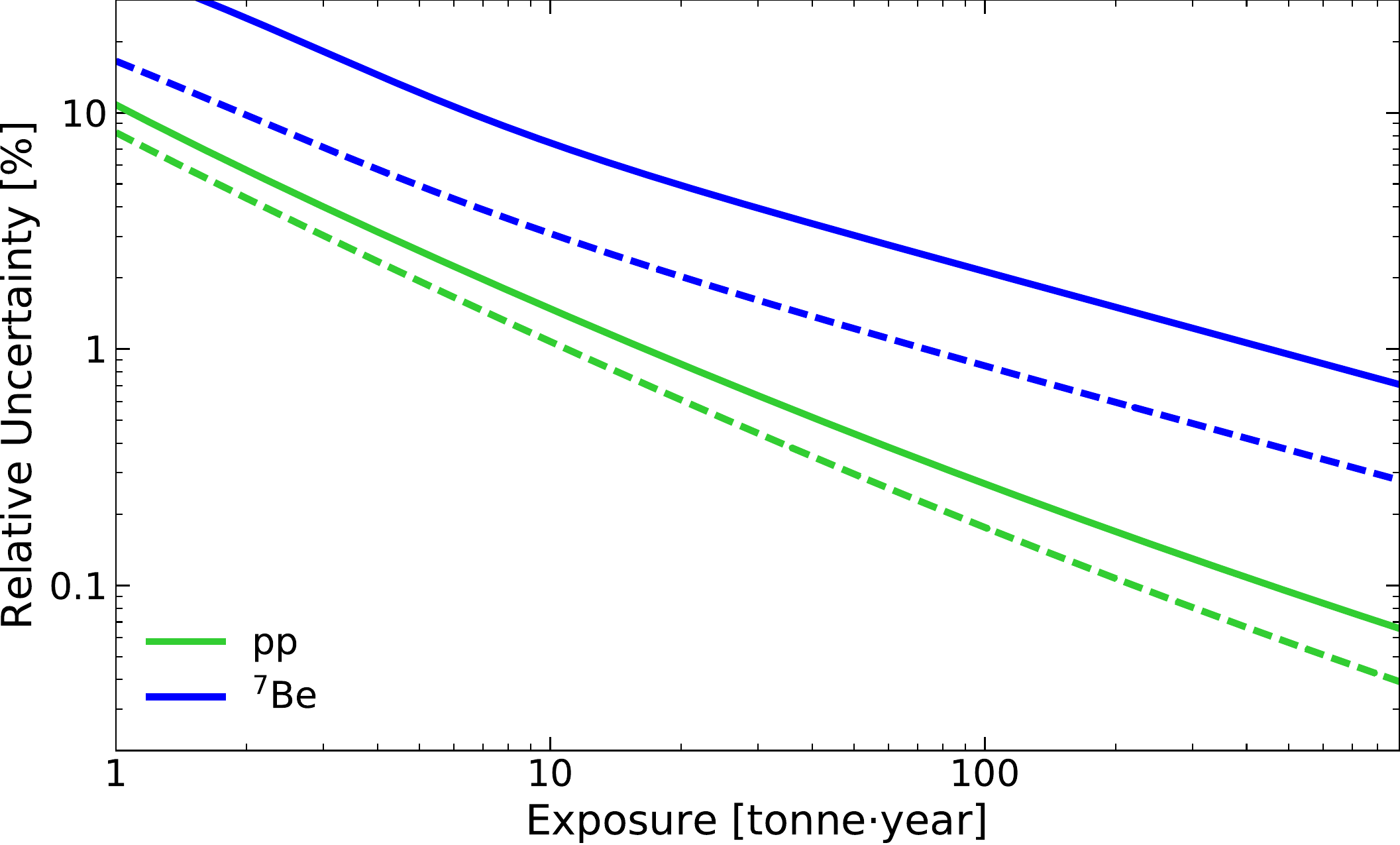}
\caption{The measured relative uncertainty of each solar neutrino component and neutrino capture as a function of exposure. The median fluxes of the high-Z model are assumed. Solid (dashed) curves correspond to a natural (depleted) target. A log scale of the $pp$ and $^7$Be components is shown in the bottom panel for clarity.}
\label{fig:precision}
\end{figure}

Toy experiments are run for each exposure in the range of interest, [1,1000]\,ty, to ascertain the expected relative 1$\sigma$ uncertainties, $\sigma_i$, for each neutrino flux component. These uncertainties are shown in Figure~\ref{fig:precision} normalized to their respective median high-Z flux values. The solid lines correspond to a natural target, while the dashed lines indicate a target depleted of $^{136}$Xe by two orders of magnitude. 

With 1\,ty, DARWIN would quickly match the precision of the $pp$ flux (10\%) currently set by Borexino. A subpercent measurement would follow with 20\,ty, ultimately reaching 0.15\% at 300\,ty. Similarly, DARWIN would match Borexino's $^7$Be measurement (2.7\%) within 60\,ty and then achieve 1\% precision with 300\,ty. The $^{13}$N and $^{15}$O neutrinos would also be attainable as independent measurements. The former (latter) would require 100\,ty (200\,ty) to reach 3$\sigma$ detection with a natural xenon target. Finally, DARWIN could observe the $pep$ component and neutrino capture process with 60\,ty and 200\,ty, respectively, using a necessarily depleted target.

The solar luminosity inferred from solar neutrino data, $L_{\odot,\nu}/L_\odot=1.04^{+0.07}_{-0.08}$, agrees with the measured (photon-inferred) solar luminosity within 7\%~\cite{Bergstrom:2016cbh}. The $pp$ reaction contributes most strongly to the total energy generation in the Sun. Thus, high-precision measurements of the $pp$ and $^7$Be components, which respectively comprise 92\% and 7.4\% of the solar lumonisity, would reduce this uncertainty. With the precision levels shown here, DARWIN would achieve an uncertainty of 0.2\% on the neutrino-inferred solar luminosity.

%% Particle parameter measurement DARWIN could make

\begin{figure}
\centering
\includegraphics[width=1\columnwidth]{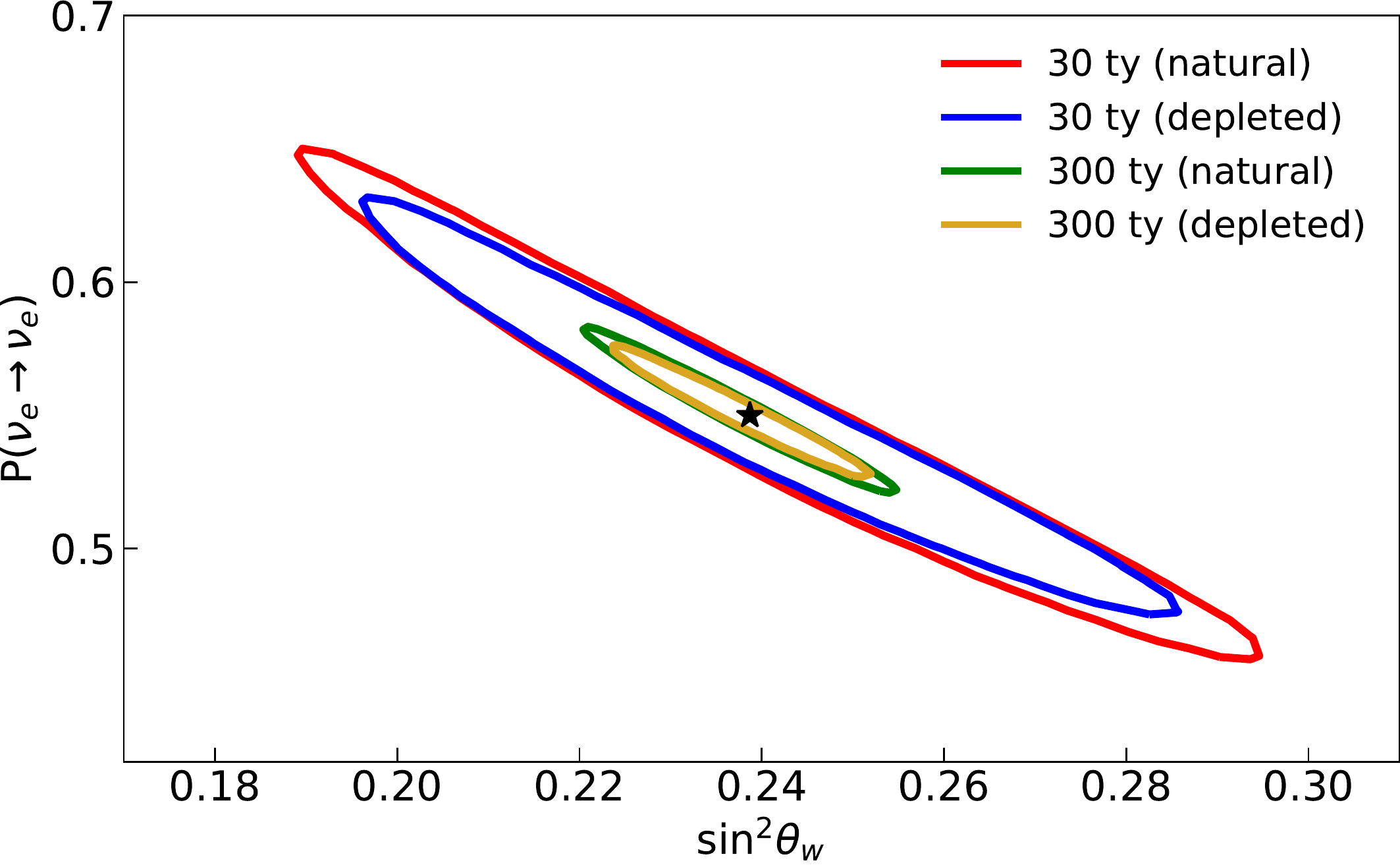} \\
\includegraphics[width=1\columnwidth]{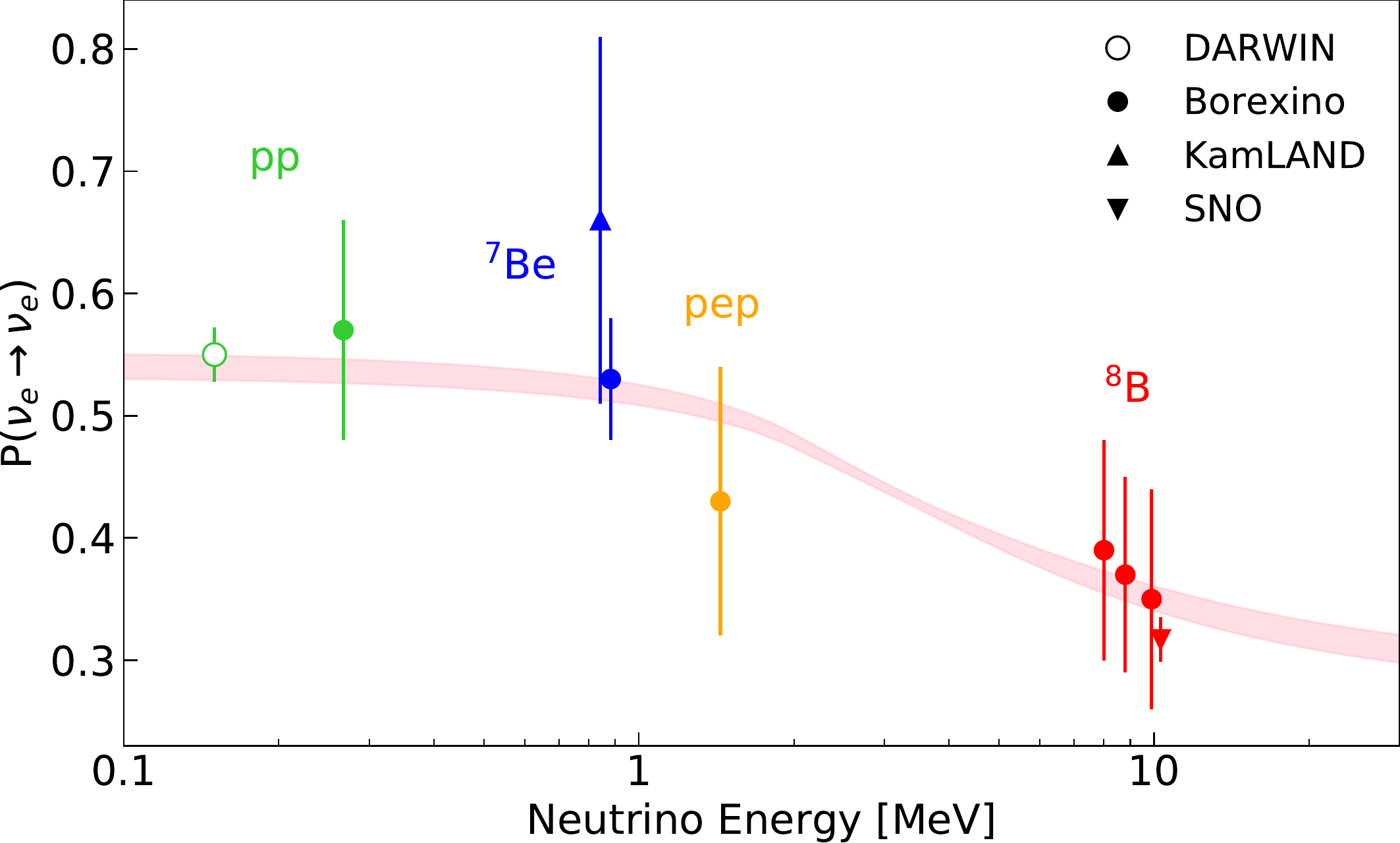} \\
\caption{(top) The 68\% confidence regions of $\sin^2\theta_w$ and $P_{ee}$ for two exposures and the two target compositions. (bottom) The $\nu_e$ survival probability versus neutrino energy under the high-Z SSM. Dots represent the solar measurements of $pp$ (green), $^7$Be (blue), $pep$ (orange), and $^8$B (red) from Borexino~\cite{Agostini:2018glp,Agostini:2017cav}. The upward (downward) triangle shows a measurement of $^7$Be ($^8$B) from KamLAND (SNO)~\cite{Abe:2011em,Aharmim:2011vm}. The open point indicates DARWIN's projected enhancement of the precision of the $\nu_e$ survival probability to 0.02 below 420\,keV using $pp$ events. The pink band represents the 1$\sigma$ prediction of the MSW-LMA solution~\cite{Capozzi:2017ipn}.}
\label{fig:results}
\end{figure}

\section{Electroweak and Oscillation Parameters}
Following a precise measurement of the $pp$ component, one may infer the values of the electroweak mixing angle and the $\nu_e$ survival probability, as they directly affect the shape of its observed recoil spectrum. A likelihood function in which the electroweak and oscillation parameters are free to vary, while the flux scales (see Table~\ref{table:solpars}) remain fixed, is adopted:
\begin{linenomath}\begin{equation}
P(n_j | \mu_j(\theta)) = \mathcal{L}(\theta) .
\end{equation}\end{linenomath}
The presence of $^7$Be neutrinos only slightly worsens the sensitivity to these parameters, while the other neutrino components have a negligible effect. The uncertainty in the $pp$ flux contributes negligibly to the total uncertainty of $\sin^2\theta_w$ and $P_{ee}$.

One finds the maximum likelihood estimators of $\sin^2\theta_w$ and $P_{ee}$ in a series of toy experiments. From the resultant 2D distribution, the 68\% confidence regions are determined, as shown in Figure~\ref{fig:results}~(top), for four scenarios based on two exposures (30 and 300\,ty) and two target compositions.

In the case of a natural target, DARWIN would reconstruct $\sin^2\theta_w$ and $P_{ee}$ with uncertainties as small as 0.0122 (5.1\%) and 0.022 (4.0\%), respectively. Alternatively, with a depleted target, the uncertainties would shrink to 0.0099 (4.2\%) and 0.017 (3.1\%). A measurement of $\sin^2\theta_w$ would be the first in this energy range, albeit with an uncertainty roughly five times higher than those at higher energies. A measurement of $P_{ee}$ would improve upon the existing one from Borexino by an order of magnitude. This projection is shown in Figure~\ref{fig:results}~(bottom) with solar neutrino measurements from Borexino~\cite{Agostini:2018glp,Agostini:2017cav}, KamLAND~\cite{Abe:2011em}, and SNO~\cite{Aharmim:2011vm}.

\section{Solar Abundance Problem}
DARWIN may utilize a combination of neutrino flux measurements to probe the metallicity of the Sun. The sensitivity assessment of the flux components, $\vec{f}$, described in Section~\ref{sec:fluxsen} is repeated with the same backgrounds. In this instance, however, the median flux values of the high- and low-Z models are allowed to vary according to their respective theoretical uncertainties. These uncertainties, $\sigma_i$, are then put into a multivariate (Gaussian) simulation characterized by a 5-dimensional matrix $\Sigma=\begin{bmatrix}\rho_{ij}\sigma_i\sigma_j \end{bmatrix}$ that accounts for all correlations of the flux components, $\rho_{ij}$. The correlation values are based on~\cite{Vinyoles:2016djt}. For each trial at a given exposure, the randomly sampled $\vec{f}$ is used to calculate a (squared) Mahalanobis distance $\delta^2=(\vec{f}-1)^T\cdot\Sigma^{-1}\cdot(\vec{f}-1)$~\cite{Maha:1936pni}, which quantifies the deviation from the true (median) values. The simulation is repeated for both the high- and low-Z models. A p-value is calculated for each iteration of the high-Z simulation by integrating the low-Z Mahalanobis distribution above the high-Z Mahalanobis value. The median p-value is then taken from its final distribution.

\begin{figure}
\centering
\includegraphics[width=1\columnwidth]{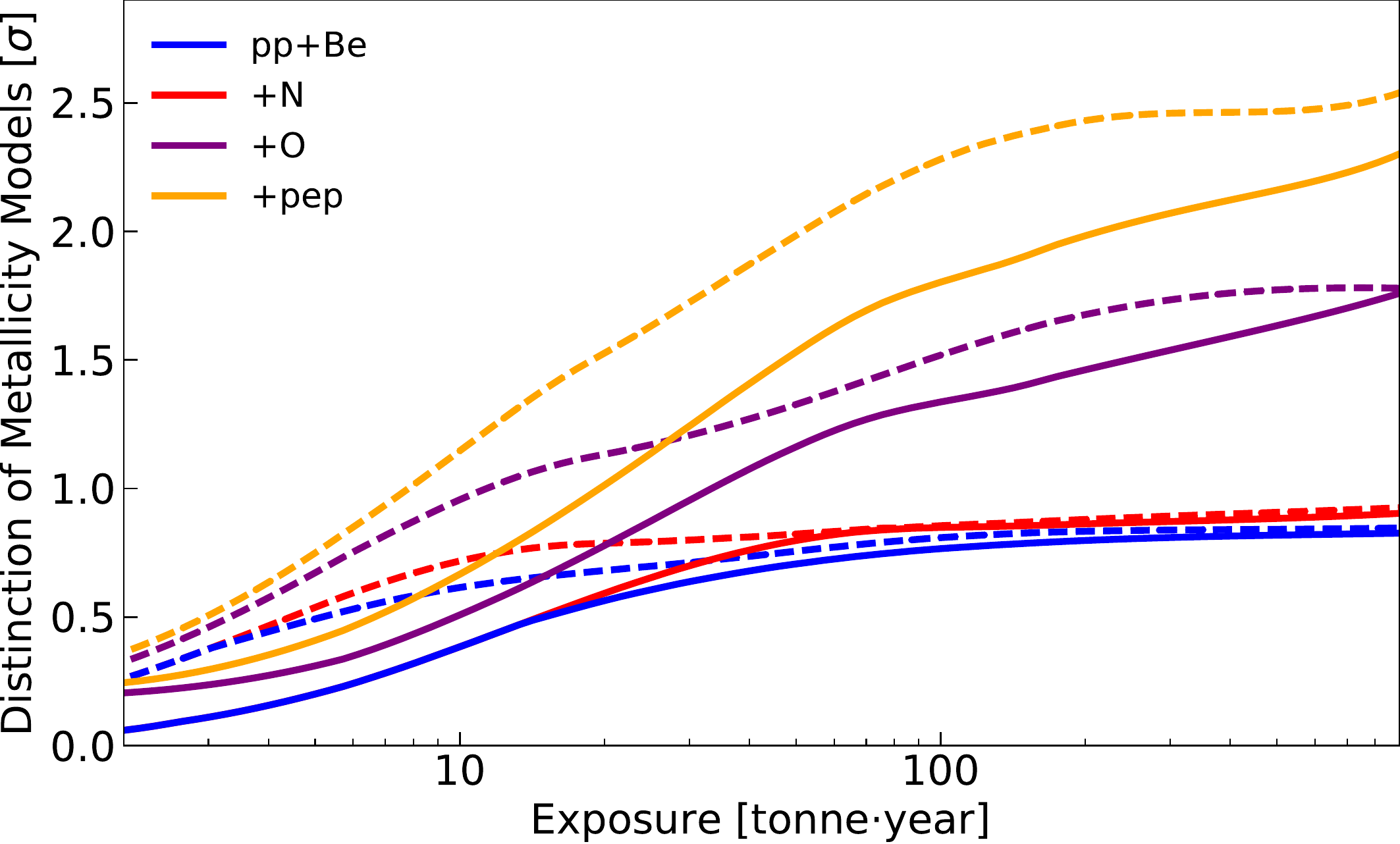}
\caption{The median significance with which the the high- and low-Z models may be distinguished
%The p-value of the low-Z model
is calculated for the first pair of flux measurements, $pp$ and $^7$Be, as a function of exposure. Additional cases add $^{13}$N, $^{15}$O and $pep$ sequentially. The solid (dashed) curves correspond to a natural (depleted) xenon target.
}
\label{fig:metallicity}
\end{figure}

Figure~\ref{fig:metallicity} displays the significance corresponding to these p-values as a function of exposure for the first pair of measurements, $pp$ and $^7$Be, and for each subsequent addition of the other components. The $^{13}$N component only modestly increases the exclusion potential due to the large uncertainties in both theory and experiment. The combination of both $^{13}$N and $^{15}$O, however, yields a significant gain above $\sim$10\,ty. The $pep$ component enhances the exclusion to 2.1$\sigma$ (2.5$\sigma$) with a natural (depleted) target at an exposure of 300\,ty. The series of improvements in the distinction power with each new measurement does not necessarily reflect the intrinsic properties of the corresponding neutrino component or its correlation to solar metallicity. Rather, it represents DARWIN's specific ability to discern one model from another using up to five measurements simultaneously, at a given exposure.

%The theoretical uncertainties define asymptotic limits for a given combination of fluxes.
DARWIN would remain limited by the $^{136}$Xe background with a natural target, but with depletion it would distinguish between the high- and low-Z SSMs up to the theoretical uncertainties. The significance illustrated in Figure~\ref{fig:metallicity} may be further improved either with a measurement of the solar $^8$B flux via CEvNS in DARWIN or with independent measurements from other experiments.

%% Addressing metallicity

\section{Outlook}
The DARWIN observatory will feature sensitivity to five components of the solar flux via ES. A low energy threshold of 1\,keV allows DARWIN to observe the majority of $pp$ neutrinos, which have (mostly) eluded contemporary neutrino observatories. With 300\,ty, DARWIN would be able to achieve 0.15\% precision in the $pp$ flux measurement, approximately two orders of magnitude better than the current precision from Borexino. DARWIN would improve upon existing measurements of the $^7$Be flux by a factor of 3. These measurements, in turn, would reduce the uncertainty on the neutrino-inferred solar luminosity to 0.2\%. The $pep$ neutrinos may be observed with 3$\sigma$ significance within the lifetime of the experiment, depending on the target composition. And, with only three years of data, DARWIN would make independent observations of CNO neutrinos with 3$\sigma$ significance.

Precise measurements of these solar components further extend the physics reach of DARWIN. The high-statistics $pp$ events would provide the means to measure both $\sin^2\theta_w$ and $P_{ee}$ in an energy region that is yet to be probed. The precision of $P_{ee}$, in particular, would be up to one order of magnitude better than the current lowest-energy measurement from Borexino. All obtained measurements and limits on the fluxes would together provide information to distinguish between the high- and low-Z SSMs. These capabilities are dependent on the target composition. Only with a target depleted of $^{136}$Xe by approximately two orders of magnitude would DARWIN make such precise measurements via ES or exploit them to distinguish between solar models. DARWIN may further enhance its distinction power with a measurement of the $^8$B flux via CEvNS. The forecast for such a measurement is highly sensitive to the achieved energy threshold for nuclear recoils, and it is left for a future study. A powerful physics case exists for the pursuit of solar neutrinos in DARWIN, and it comes without the need for additional investment beyond the option of depletion.

Data published in this article are available in~\cite{repo:2020}.

\section*{Acknowledgements}
\normalsize{
This work was supported by the Swiss National Science Foundation under grants No 200020-162501, and No 200020-175863, by the European Union’s Horizon 2020 research and innovation programme under the Marie Sklodowska-Curie grant agreements No 674896, No 690575, and No 691164, by the European Research Council (ERC) grant agreements No 742789 (Xenoscope), and No 724320 (ULTIMATE), by the Max-Planck-Gesellschaft, by the Deutsche Forschungsgemeinschaft (DFG) under GRK-2149, by the US National Science Foundation (NSF) grants No 1719271, and No 1940209, by the Portuguese FCT, by the Netherlands Organisation for Scientific Research (NWO), by the Ministry of Education, Science and Technological Development of the Republic of Serbia and by grant ST/N000838/1 from Science and Technology Facilities Council (UK).
}

\bibliographystyle{apsrev4-1}

\end{document}